The Open University of Israel

Department of Mathematics and Computer Science

# Optimizations of Management Algorithms for Multi-Level Memory Hierarchy

Thesis submitted as partial fulfillment of the requirements

towards an M.Sc. Degree in Computer Science

The Open University of Israel

By

**Gal Oren**

Prepared under the supervision of Dr. Leonid Barenboim

and under the scientific guidance of Dr. Lior Amar

December 2015

## Acknowledgments

I would like to express my sincere gratitude to my scientific supervisor over the last two years, Dr. Leonid Barenboim, and to my scientific mentor over the last four years, Dr. Lior Amar, who led me to my current achievements in the field of computer science - in theory as well as in practice.

Also, I would like to thank my research facility – Nuclear Research Center - Negev (NRCN) – and its' executives for giving me the opportunity to learn for this Master's Degree as a permanent part of my professional experience.

Finally, I would like to thank my dear friend Noam Shimoni for his editorial work of this thesis.

# Content

## List of Figures

## List of Tables

## List of Equations

**Abstract**

Memory and storage are often assumed to be unsophisticated, flat resources, with simple properties, such as a constant access time. Over the years this assumption has been proven to be wrong, and understanding of the memory hierarchy could be useful in order to enhance the performance of an algorithm or a data structure. For example, the Storage Class Memory (SCM), widely known as Persistent Memory, is a new technology which represents a new hybrid form of storage and memory with unique characteristics, meaning a memory which is non-volatile, cheap in per bit cost, has fast access times for both read and writes, and is solid state.

Operating Systems are likely use the SCM as either very fast block storage devices formatted by file systems and databases, or as direct memory mapped "files" for the next generation of programs. In the near future the SCM is predicted to modify the form of new programs, the access form to storage, and the way that storage devices themselves are built. Therefore, a combination between the SCM and a designated Memory Allocation Manager (MAM) that will allow the programmer to *manually* control the different memories in the memory hierarchy will be likely to achieve a new level of performance for memory-aware data structures. Although the manual MAM seems to be the optimal approach for multi-level memory hierarchy management, this technique is still very far from being realistic, and the chances that it would be implemented in *current* codes using High Performance Computing (HPC) platforms is quite low.

This premise means that the most reasonable way to introduce the SCM into any usable and popular memory system would be by implementing an *automated* version of the MAM using the fundamentals of paging algorithms, as used for two-level memory hierarchy. Our hypothesis is that achieving appropriate transferability between memory levels may be possible using ideas of algorithms employed in current virtual memory systems, and that the adaptation of those algorithms from a two-level memory hierarchy to an N-level memory hierarchy is possible.

In order to reach the conclusion that our hypothesis is correct, we investigated various paging algorithms, and found the ones that could be adapted successfully from two-level memory hierarchy to an N-level memory hierarchy. We discovered that using an adaptation of the Aging paging algorithm to an N-level memory hierarchy results in the best performances in terms of Hit / Miss ratio. In order to verify our hypothesis we build a simulator called "DeMemory simulator" for analyzing our algorithms as well as for other algorithms that will be devised in the future.

# 1 Data Structure Algorithms Implementation in Memory Hierarchies

## 1.1 Introduction: Memory Hierarchy Awareness

Often memory is assumed to be an unsophisticated, flat resource, with simple properties, much like a constant access time [1]. Generally speaking, this is seldom the case, because common computers often have five memory layers with different properties. Three of these layers dwell on the processor chip, one layer is the RAM memory and one layer is the physical memory, such as SSD and HDD (Figure 1). The layers on the processor chip are referred to as L1 cache, L2 cache and L3 cache. L1 cache is a rather small piece of memory with extremely high access time, used directly by the processor. L2 cache is slightly slower and vastly larger than L1 cache. L3 cache is slower than L2 and L1 caches, and it is shared by all cores. When accessing memory, the CPU will look in the main memory only after it looks in L1 cache, L2 cache, and L3 cache. In the following chapters we will refer to this memory hierarchy using the term "two-level memory hierarchy", excluding the three cache levels from our discussion, because of our sole interest in the RAM-HDD match.

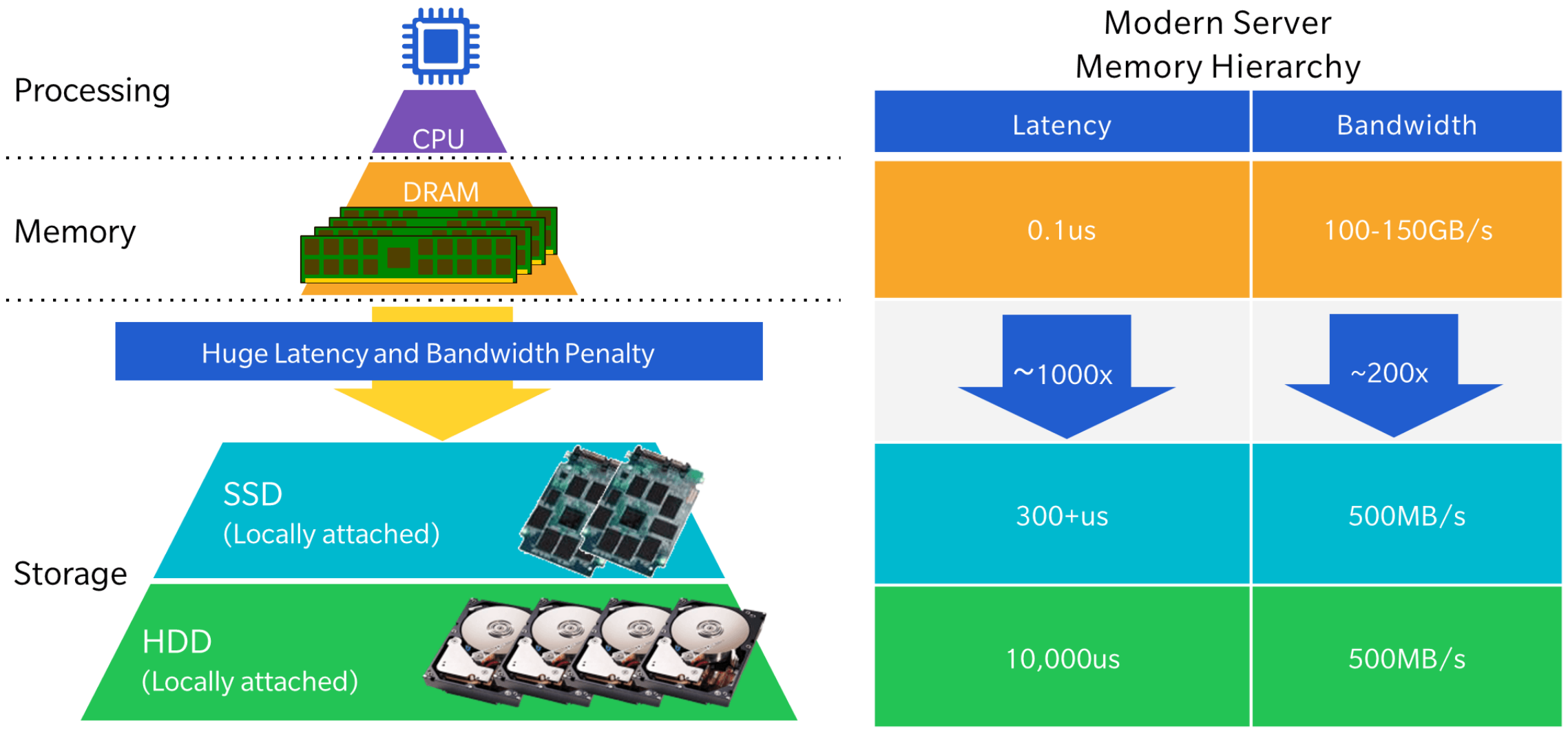

Figure 1: **The Memory Hierarchy and its Latency and Bandwidth Parameters** (Rambus, 2015). The latency and bandwidth penalty between the two upper levels and the two lower levels is about 2-3 orders of magnitude.

Understanding the memory hierarchy could be useful in order to enhance the performance of an algorithm or a data structure [2]. Algorithms and data structures that adjust to a specific memory organization are known as memory-aware or memory-conscious. Algorithms and data structures that are planned to execute well with memory unawareness, independently of the memory parameters and hierarchy are called memory-oblivious. Design of memory-aware algorithms requires awareness of the memory hierarchy.

## 1.2 Previous Work: Usage of the Memory – Paging vs. STXXL

Operating Systems (OS) implement the virtual memory mechanism that extends the working space for applications, mapping an external memory file (page/swap file) to virtual addresses. This idea supports the Random Access Machine model [3] in which a program has an infinitely large main memory. With virtual memory, the application does not know where its data is located, whether in the main memory or in the swap file. This abstraction does not have large running time penalties for simple sequential access patterns: The OS is even able to predict them and to load the data ahead of time.

For more complicated patterns these remedies are not useful and even counterproductive: The swap file is accessed very frequently; the executable code can be swapped out in favor of unnecessary data; the swap file is highly fragmented and thus many random I/O operations are needed even for scanning.

The OS cannot adapt to complicated access patterns of applications dealing with massive data sets. Therefore, there is a need for explicit handling of external memory accesses. The applications and their underlying algorithms and data structures should care about the pattern and the number of external memory accesses (I/Os), which they cause.

Several simple models have been introduced for designing I/O-efficient algorithms and data structures (also called *external memory* algorithms and data structures). The most popular and realistic model is the Parallel Disk Model (PDM) of Vitter and Shriver [4]. In this model, I/Os are handled explicitly by the application.

An I/O operation transfers a block of $B$ consecutive elements from/to a disk to minimize the latency. The application tries to transfer $D$ blocks between the main memory of size $M$ bytes and $D$ independent disks in one I/O step to improve bandwidth (Figure 2). The input size is $N$ bytes which is quite larger than $M$.

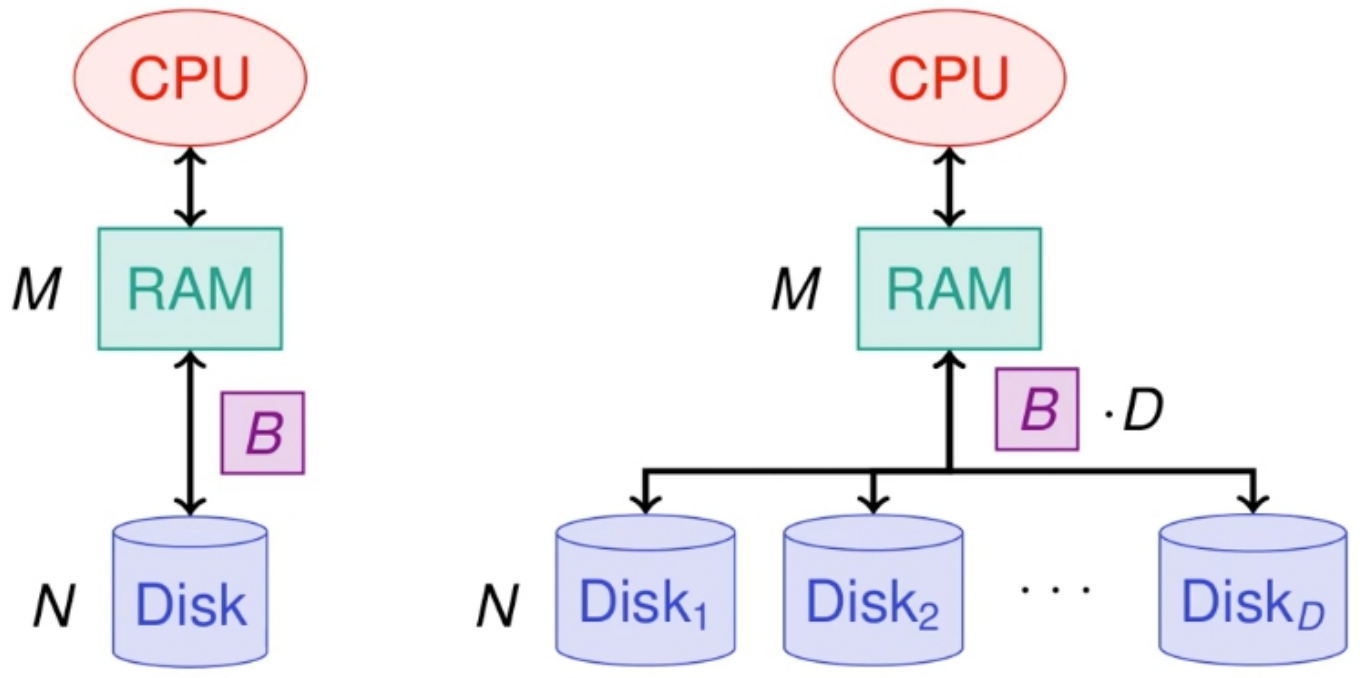

Figure 2: **Serial Disk Model (SDM) vs. Parallel Disk Model (PDM)** ([5]).

The most common implementation of the PDM model can be found at the STXXL project [5]. The core of STXXL is an implementation of the C++ standard template library STL for *external memory* (out-of-core) computations, i.e., STXXL implements containers and algorithms that can process huge volumes of data that only fit on disks. While the compatibility to the STL supports ease of use and compatibility with existing applications, another design priority is high performance. The performance features of STXXL include: Transparent support of multiple disks, variable block length, overlapping of I/O and computation, and prevention of OS file buffering overhead.

## 1.3 Future Memory: Storage Class Memory

Storage is considered to be a mechanical HDD that supplies virtually unlimited capacity, when compared to DRAM, and it is also perpetual, which means that data is not lost if the computer happens to crash or disconnect from electricity. The issue with hard drives is that in various situations they are unable to supply data to applications with the sufficient speed [6].

Storage Class Memory (SCM) proposes to minimize or even close the widening gap between CPU processing speeds, the need to rapidly transfer big data blocks, and the read-write speeds suggested by HDD reliant systems. The SCM [7], widely known as Persistent

Memory, is a technology which represents a new hybrid form of storage and memory with unique characteristics, meaning a memory which is non-volatile, cheap in a per bit cost, has fast access times for both read and writes, and is solid state.

The SCM has a unique mechanism. Created out of flash-based NAND, SCM is a new form of storage that can provide a middle step between high-performance DRAM and cost-effective HDDs (Figure 3). It may very well provide read performance analogous to DRAM (perhaps even better in some cases), and write performances that are significantly faster than HDD technology (factors of hundreds better than HDD and even beyond). Also, it is predicted that the production costs of SCM and HDD will be broadly similar by the end of this decade [8].

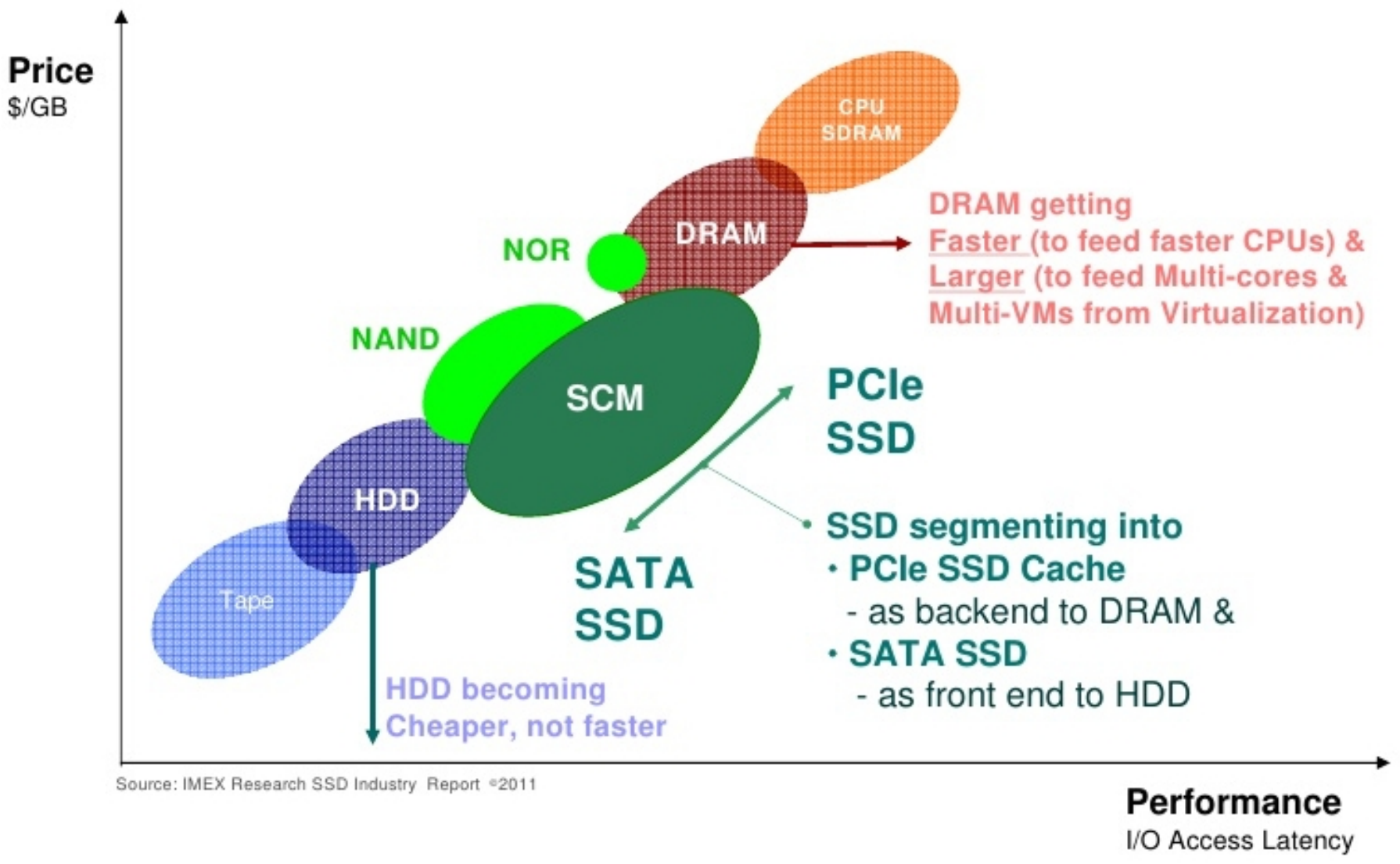

Figure 3: **The SCM in Context of the Available Memories** (IMEX Research, 2011).

Those new SCM storage-memory systems connect to memory slots in a server and are mapped and accessed in the same fashion as the memory, even though they are slightly slower, and they can be addressed atomically at either the byte or the block level, unlike previous eras of storage technology. The SCM can be used directly as execution memory or data storage memory. The current SCM products include the improved Flash [9], the Phase Change Memory (PCM) [10], the Magnetic RAM (MRAM) [11], the Solid Electrolyte RAM – Nano-Ionic RAM [12], the Ferroelectric RAM (FRAM) [13] and the Memristor [14].

Operating Systems are likely to use the SCM as either very fast block storage devices formatted by file systems and databases, or as direct memory mapped “files” for next generation of programs. In the near future the SCM is predicted to modify the form of programs, the access form to storage, and the way that storage devices themselves are built. Therefore, a combination between SCM and the existing memories, using a new memory allocation manager that will act like STXXL, will be likely to achieve a new level of performance for memory-aware data structures. Such a memory system with different kinds of memory speeds, access fashions and volumes is modeled by a collection (sorted according to memory speeds) of N arrays, such that each array represents a memory level, where level 1 is the fastest and level N is the slowest.

## 1.4 The Memory Allocation Manager (MAM)

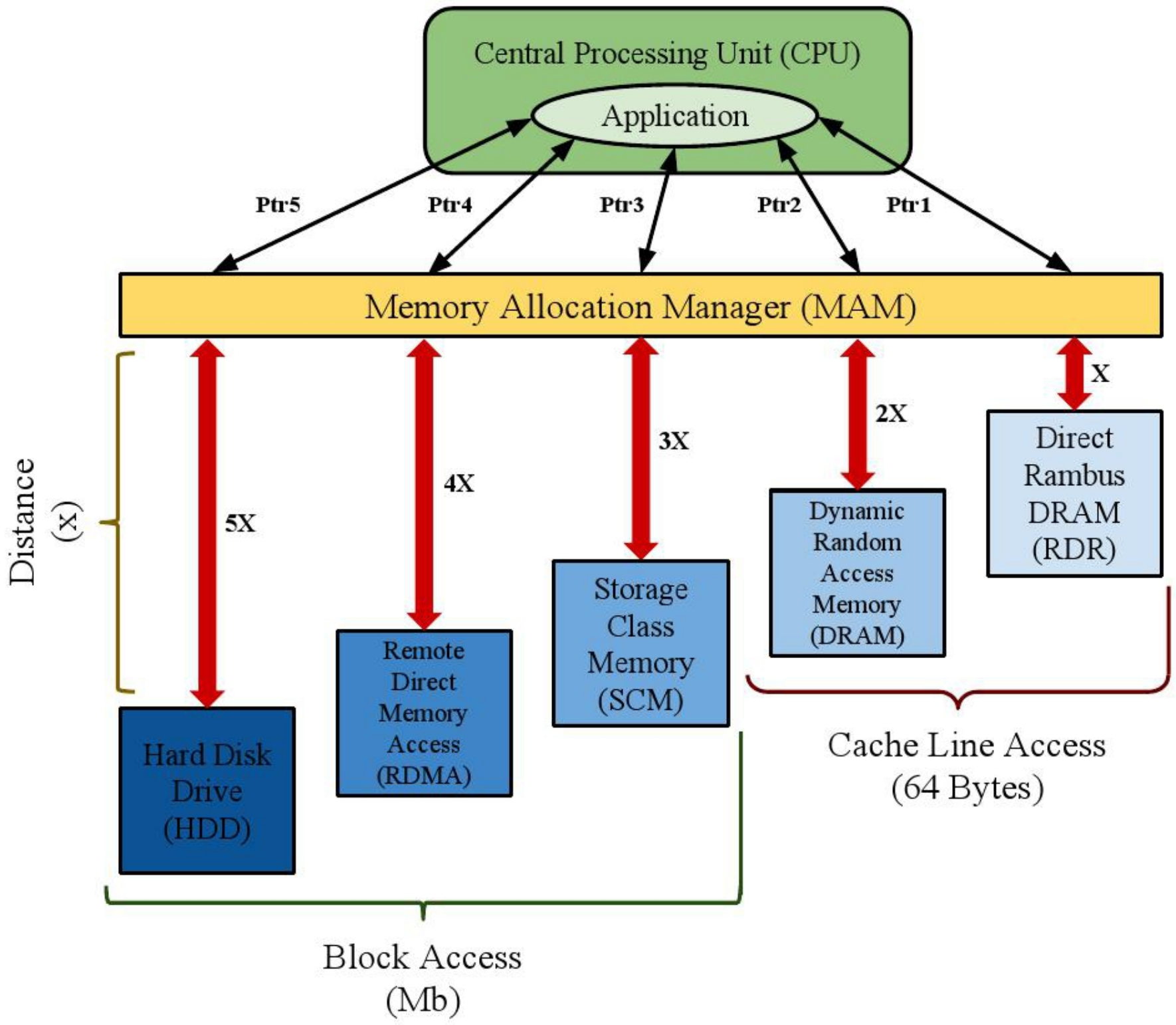

Figure 4: **The Memory Allocation Manager (MAM) Diagram of Usage.**

The Memory Allocation Manager (MAM), based on the idea of C language malloc [15] and STXXL allocation modules, will take the power of the memory management from the virtual memory mechanism and will deliver it to the programmer. In this scenario there is no common dichotomy between a single memory and a single storage, but a plurality of memories, which are the new MAM, by the instructions from the application, will take charge of them.

The programmer will have the option to control those memories – serially or in parallel – in order to achieve the best performance possible.

*void *ptr1, *ptr2*

*ptr1 = alloc_cache_mem(...) ; free_cache_mem(ptr1)*

*ptr2 = alloc_block_mem(...) ; free_block_mem(ptr2)*

In each of those categories the programmer will be able to set three elements:

1. The amount of memory the application needs (100MB, 1TB etc.) – Mandatory.
2. The speed degree of the memory (Fast, Medium, Slow) which depends on two factors:
   2.1. The physical distance between the CPU and the memory.
   2.2. The type of memory (RDR vs. DRAM in the cache line access category or Storage Class Memory vs. HDD in the block access memory category).
3. A pointer to an opposite type of allocated memory, which will create a synchronization between the current allocation and the previous one in a paging style fashion – Optional.

*void *ptr1, *ptr2*

*ptr1 = alloc_block_mem(1TB, Slow, ptr2) | free_block_mem(ptr1)*

*ptr2 = alloc_cache_mem(100MB, Fast, ptr1) | free_cache_mem(ptr2)*

First, a data structure (hash table, in this example) will be created on the chosen memory, and it will be specified how much memory should this data structure use from its main memory type.

*slow_ht = ht_create(500GB, ptr1)*

*fast_ht = ht_create(50MB, ptr2)*

After the initiation of the cache line access memories and the block access memories it will be possible to use the different memories.

- *Insert:* Insertion of a key-value arguments to a specific type of memory.
- *Get:* Get a hash value of specific key from a specific type of memory.
- *Remove*: Remove a specific value of specific key from a specific type of memory.

*Insert(key, value, Fast_HT)*

*Get("Don Draper", Slow_HT)*

*Remove("Dick Whitman", Fast_HT)*

At the end of use, the allocated hash tables memories should free themselves by a *free* command. Note, that this is not a de-allocation of the main memories, but just of the hash table that is allocated on it.

*Free(Fast_HT, Slow_HT)*

### 1.5 Data Structures Usage of the Memory Allocation Manager

There are clear benefits of using the Memory Allocation Manager (MAM) API at the design and implementation stage of any data structure, which needs to handle massive data sets. For example, here are five possible options to use the MAM to optimize hash tables performance.

1. Given that the keys are arranged in a list form, there is an option that the first immediate key will be stored in the fastest memory available, and as far as the list expands, the other keys will be stored in slower memories.
2. Using the parallel fashion of the memory manager, there is an option to check for a match to the hash function result in all the different memories at once, and by that to create a parallel access to the different serial memories.
3. Start filling the fastest memory with the hash function results, and only when it reaches its capacity continue to the slower memory. This memory allocation fashion can be repeated to the slowest memory available.
4. During the usage of the hash table, the data structure will create a dynamic histogram of the most and the least called keys and by that will rearrange their locations in the memories, from the fastest to the slowest.
5. When the keys are already known, the data structure will create a histogram of the keys, and will divide the most common keys to the fastest memories and the rarest keys to the slowest memories.

Therefore, it seems reasonable to re-modify the data structures to use those kinds of paradigms using the MAM. **However, this kind of solution arises a problem, which lies within the solution**. The reason for this problem is that the algorithms require not only to re-modify the memory platform and the access to it, but also the code that is currently based on two-level memory hierarchy. Thus a practical resort, which will represent a compromise, is in need.

## 1.6 Paging Algorithms as a Practical Resort

Although the Memory Allocation Manager seems to be the best approach for multi-level memory hierarchy management, this technique is still very far from being realistic, and the chances that it would be implemented in current codes using High Performance Computing (HPC) platforms is quite low. Considering the following reasons, it seems that the MAM will only stay on paper (or in code editors) and will not be used in most of current and near future systems. The reasons are:

1. A rewrite of source codes in a way that will be suitable to work with the MAM is not realistic, specifically and especially for HPC usages. Most of the research centers which use large computer clusters have programs that used to be operational for ages and are very big, complex and sustainable. Hence, the solution of rewriting the whole code and redefining the data structures to use the MAM is not possible, and can be efficient only for several specific application that are written today, yet not for past applications.
2. Because Storage Class Memory (SCM) is breakthrough technology, it would not be reasonable to wait for several more decades and see if people will change their codes. Instead, upgrading HPC clusters to a new level of sophistication and performance without any need to rewrite the code or manage the memory would be wiser.
3. Even the previous platforms, such as STXXL, did not become widespread among developers, although those platforms were introduced in times when memory layers were actually more expensive than current times. This indicates that using MAM would not be beneficial today for most purposes.

Therefore, an implementation of a multi-level memory hierarchy management should not incur any changes in the data structures. This premise means that the most reasonable way to introduce the SCM into any usable and popular memory system would be by implementing an *automated* and *not explicit* version of the MAM. Thus, introducing the MAM concepts and the multiple memory levels awareness into the classic paging algorithms can be a good solution, which will not be a big compromise. This idea can be achieved by introducing a new automated layer that will select the most appropriate memory levels for allocating space, and that will move data between memory levels for optimizing performance in the fashion of paging algorithms.

# 2 Page Replacement Algorithms Implementation in Memory Hierarchies

## 2.1 Usage of Paging Concept in Memory Hierarchies

As mentioned in the previous chapter, there are clear benefits in using the Memory Allocation Manager (MAM) API at the design and implementation stage of any data structure which needs to handle massive data sets, especially hash tables. Also, we offered five conceivable options to use it in order to optimize those hash tables performance. Among those five options, the most reasonable and logical option for optimization, which also matches the current paradigm in a two-level memory hierarchy, is the one in which the following hold: It is given that the keys are arranged in a list form; There is an option that the first immediate key will be stored in the fastest memory available; And as far the list expands, the other keys will be stored in the slower memories.

However, for supporting that kind of optimization, the MAM alone is not sufficient because the hash table implementation requires a collision resolution policy. This means that in some point there is going to be a shortage of memory space and there will be a need to make transfers between the different kinds of memory, from the faster ones to the slower ones, or the opposite. We also argued that transformation of the code of most current High Performance Computing (HPC) services for the MAM architecture is not realistic. Therefore, we propose to employ an additional automated and not code explicit layer that determines the appropriate levels in the memory hierarchy for placing data. This layer receives memory allocation requests (without level specifications), and determines the appropriate memory level in which allocations will be made. In addition, this layer moves data between the different memory levels in order to optimize performance.

Our hypothesis is that achieving that kind of transferability between memory levels may be possible using ideas of algorithms employed in current virtual memory system, and that the adaptation of those algorithms from a two-level memory hierarchy to an N-level memory hierarchy may be possible. In that notion, each virtual memory page is a hash table entry (or entries), and each of the virtual memory swaps can be done like the MAM would do if it would implemented with knowledge of the appropriate levels, regardless of the access fashion.

In order to reach a conclusion that our hypothesis is correct, we needed to clarify which of the paging algorithms could be adapted successfully from a two-level memory hierarchy to an N-level memory hierarchy. For doing so, we have to thoroughly investigate the current paging mechanism and the main paging algorithms.

## 2.2 Basic Paging Mechanism

It is widely known [16] [17] [18], that an Operating System has to select a page to evict from memory to create a space for the page that has to be inserted in, when a page fault happens. If the page has not been altered, the copy on the hard drive is already up to date, so no rewrite is necessary, and the page to be read in just overwrites the page being removed. If, however, the page to be removed has been modified while in memory, it must be rewritten to the hard drive to keep the hard drive copy updated.

Although it is possible to select a random page to be removed at each page fault, system performance would benefit greatly if a page that is not massively used is picked up. If a massively used page is evicted, it will apparently have to be brought back in soon, resulting in unnecessary overhead. A lot of scientific research has been performed on the subject of page replacement algorithms, both theoretical and experimental.

## 2.3 The Optimal Page Replacement Algorithm

The optimal prospective page replacement algorithm is simple to describe but cannot be implemented, thus there is no expectancy that usage of a simply described paging technique in N-level memory hierarchy will match the theoretical results.

The optimal page replacement algorithm is the one in which the following hold: When a page fault happens, some series of pages is in the memory system. One of these pages will be referenced on the upcoming instruction. The pages that have not been referenced may be non-referenced even after unlimited amount of instructions later. Each page can be tagged with the amount of instructions that will be executed prior to the time that that page is first referenced. Simply, the optimal page algorithm evicts the page with the highest tag.

The only problem with the optimal algorithm is that it is unimplementable, because at the moment of a page fault the Operating System (OS) has no information about the time when each of the pages will be referenced next. But, using a simulator which will keep track of all page references, it is possible to execute an optimal page replacement on the *second* run by using the page reference database accumulated during the *first* run.

Using this method it is possible to compare the performance of implementable algorithms with the best possible one. If for instance an OS accomplishes achieving a performance of about 10% worse than the optimal algorithm, effort spent in searching for a better algorithm will produce at most a 10% improvement. To prevent possible confusion, it should be stated that this database of page references refers only to the one simulation that has just been measured and then with only one particular input. The page replacement algorithm derived from it is therefore specific to that one simulation and input data. Of course, despite the fact that this method is useful for evaluating page replacement algorithms, it has no usage in practical systems. In the sequel we will present algorithms that *are* useful on real systems, either with two-level or N-level memory hierarchies.

It is important to mention that in hash table insertions we sometimes can predict the optimal page replacement at the *first* run using an analysis on the entries themselves, but it is not always guaranteed that that kind of analysis will be possible or plausible. Therefore, the comparison method between the first and the second run will probably be more effective to discover which is the optimal paging algorithm who suits best to the specific characterization of the insertion to the hash table. But, because this is not practical for real systems we will present modifications to the current two-level memory page replacement algorithms and will transform them to N-level algorithms.

## 2.4 The N-Level Not Recently Used (NRU) Page Replacement Algorithm

In order to let the Operating System gather meaningful statistical information about which pages are being referenced and which ones are not, most computers with a virtual memory mechanism have two status bits bound to each page. The $R$ bit is set every time the page is referenced (either read or written). The $M$ bit is set every time the page is written to. The $R$ and $M$ bits are included in each page table entry (Figure 5), and they must be updated on every memory reference, therefore it is necessary that they are set by the hardware. Whenever a bit is set to 1, it stays in that condition until the OS resets it to 0 in software.

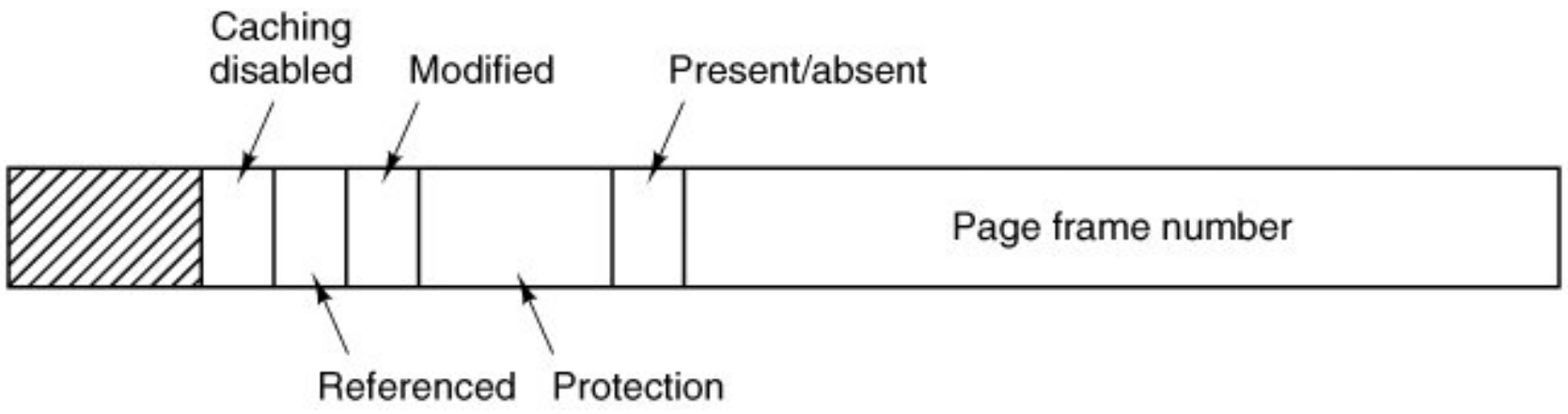

Figure 5: **The Page-Table Entry** (BSODTutorials, 2013).

The referenced and modified bits are usable to build a straightforward paging algorithm in which the following hold: When a process starts to run, the $R$ and $M$ page bits for all of its pages are set to 0 by the OS. On each clock interrupt, the operation system clears the $R$ bit, to differentiate pages that have been referenced lately from those that have been not, but do not clear the $M$ bit because this data is necessary to understand whether the page has to be rewritten to the disk or not. Whenever a page fault happens, the OS checks all the pages and splits them into 4 subsections based on the present values of their $R$ and $M$ bits, in which the following hold: Subsection 0: not referenced, not modified; Subsection 1: not referenced, modified; Subsection 2: referenced, not modified; and subsection 3: referenced, modified.

The NRU (Not Recently Used) algorithm evicts some page from the lowest numbered nonempty subsection. Therefore it is obvious that the meaning of the algorithm is that it is better to evict a modified page that has not been referenced in at least one clock interrupt than a clean page that is used massively. Thus, the central attractiveness of the NRU paging algorithm is that it is very conceivable, does not require too many resources to implement, and gives a performance that may be adequate even though not optimal.

Therefore, the transition of this algorithm to an N-level memory hierarchy model is fairly simple, because it only requires each level of memory to set a different clock tick rate. In this hypothesis, the clock tick rate needs to be proportionate to the access speed of the specific level of memory, and the pages will *"diffuse"* from one layer of memory to the other by demand. By this division of rates it is possible to get a uniformed removals of modified pages that have not been referenced over the N-level memory hierarchy (L = 1: Highest, L = N: Lowest).

<u>*Formulation of the NRU algorithm for N-level memory hierarchy*</u>:

- *Set* memory levels to N (ML = N).
- *Set* current memory level pointer to the highest (L = 1).

1. **Insertion** of a new page:
    - 1.1. *If* the current memory level pointer is higher than the lowest memory level (L > ML) :
        - 1.1.1. *Return* False. /* Recursion Termination */
    - 1.2. **Call** to the page.
    - 1.3. *If* the page exists in the memory / storage:
        - 1.3.1. *Check* if placing in the L-level of memory is possible.
        - 1.3.2. *If* placement possible:
            - 1.3.2.1. *Place* page at the L-level of memory.
            - 1.3.2.2. *Return* True. /* Recursion Termination */
        - 1.3.3. *Else If* placement impossible:

1.3.3.1. **Remove** the Not Recently Used page.

1.3.3.2. *Place* the page instead of the removed page.

1.3.3.3. *Do* **Insertion** of the Not Recently Used page to a lower level (L = L+1) /* Recursion Invocation */

1.4. *Else If* page does not exist in the memory / storage:

1.4.1. *Return* False. /* Recursion Termination */

2. **Call** to a page:

2.1. *Calculate* the addressing of the page in the memory / storage.

2.2. *If* page found:

2.2.1. *Return* Real Addressing.

2.3. *Else If* page was not found:

2.3.1. *Return* False.

3. **Remove** of a specific page:

3.1. *Store* the page in a temporary storage.

3.2. *Free* the addressing of the page.

3.3. *Return* the page from the temporary storage.

4. **Update** of an existing page (*by the OS*):

4.1. *If* Read / Write action performed on the page:

4.1.1. *Set* R bit to 1 (R = 1).

4.2. *If* Modification action performed on the page:

4.2.1. *Set* M bit to 1 (M = 1).

4.3. *If* clock interrupt:

4.3.1. *Set* R bit to 0 (R = 0).

## 2.5 The N-Level First-In-First-Out (FIFO) Page Replacement Algorithm

The First In First Out page replacement algorithm (also known as FIFO) evicts the page that aged in the Operating System (OS) for the longest period of time. The OS follows after the arrangement in which pages are placed in the main memory. When there is a need to evict a page, the algorithm selects the one that aged in the main memory for the longest period of time. Intuitively, it is plausible that the selected page for eviction has had its opportunity to be referenced and it is time to give another page that opportunity.

The problem with the first-in-first-out algorithm is that it can replace massively used pages too, which would be a very unfortunate selection, because the page would be called back to main memory almost at the same time it was evicted – a situation that will increase the page-fault rate. This poor chain of events can be controlled and eliminated by implementing FIFO with a referenced bit for each page and evicting a page only if its referenced bit is set to zero. The *second-chance* variation of FIFO do so by checking the referenced bit of the most aged page in the following way: If the R bit is equal to 0, the second-chance variation instantly chooses that page for eviction. But, if the R bit is equal to 1, the algorithm sets the bit as 0, and moves the page to the tail of the FIFO queue. By doing so, a page of this kind is handled basically in the same way as a new arrival page. Gradually, the page moves towards the head of the queue. At the time when the page arrives at the head, it will be chosen for eviction only if the referenced bit is equal to 0. Active pages – in which their R bits are still equal to 1 – will be chosen to go back to the tail of the list, and therefore they will remain in main memory.

The *clock* page replacement algorithm, which actually results in the same outcome as the second-chance page replacement algorithm, organizes the pages in a cyclic list instead of a regular list. At every occasion when a page fault happens, a list pointer revolves around the cyclic list in the same fashion as the spin of a hand of the clock. At the time a page's R bit is equal to 0, the pointer is moved to the next part of the list, imitating the transfer of this page to the back of a FIFO queue. The clock page replacement algorithm sets fresh arrivals in the first page it comes across with the R bit equals to 0.

Therefore, the adaptation of this algorithm (with the second-chance and clock page replacement strategy) to an N-level memory hierarchy model requires a *gradual* eviction of pages to a lower level of memory. When the referenced bit of the oldest page is off, the algorithm should immediately select that page for replacement. At that stage, the algorithm will need to decide to which level of memory it will be the best to evict the page. Based on the reasonable assumption that the faster the level of memory is, the better it will be to place the evicted page to; and based on the knowledge that the upper levels of memory are not an option for eviction, the algorithm will try first to transfer the selected page of a level i to the (i+1)-level of memory. In case of failure to transfer the page due to a full capacity situation in the (i+1)-level, the algorithm will try to place the page in a lower level of memory (i+2, i+3, etc.) until the eviction process succeeds.

*Formulation of the FIFO algorithm for N-level memory hierarchy:*

- *Set* memory levels to N (ML = N).
- *Set* current memory level pointer to the highest (L = 1).

1. **Insertion** of a new page:

    1.1. *If* the current memory level pointer is higher than the lowest memory level (L > ML) :

    1.1.1. *Return* False. /* Recursion Termination */

    1.2. **Call** to the page.

    1.3. *If* the page exists in the memory / storage:

    1.3.1. *Check* if placing in the L-level of memory is possible.

1.3.2. *If* placement possible:

1.3.2.1. *Place* the page at the L-level of memory.

1.3.2.2. *Return* True. /* Recursion Termination */

1.3.3. *Else If* placement impossible:

1.3.3.1. *If* the reference bit of at least one of the pages is set to 0 (R = 0):

1.3.3.1.1. **Remove** a page with reference bit set to 0.

1.3.3.1.2. *Place* the page instead of the removed page.

1.3.3.1.3. *Do* **Insertion** of the page with reference bit set to 0 to a lower level (L = L+1). /* Recursion Invocation */

1.3.3.2. *Else If* the reference bit of all of the pages is set to 1 (R = 1):

1.3.3.2.1. *Do* **Insertion** of the page to a lower level (L = L+1). /* Recursion Invocation */

1.4. *Else If* page does not exist in the memory / storage:

1.4.1. *Return* False. /* Recursion Termination */

2. **Call** to a page:

2.1. *Calculate* the addressing of the page in the memory / storage.

2.2. *If* page found:

2.2.1. *Return* Real Addressing.

2.3. *Else If* page was not found:

2.3.1. *Return* False.

3. **Remove** of a specific page:

    3.1. *Store* the page in a temporary storage.

    3.2. *Free* the addressing of the page.

    3.3. *Return* the page from the temporary storage.

4. **Update** of an existing page (*by the OS*):

    4.1. *If* Read / Write action performed on the page:

        4.1.1. Set R bit to 1 (R = 1).

    4.2. *If* the page reached to the head of the FIFO queue *and* the page reference bit is on (R = 1):

        4.2.1. Turn off the reference bit (R = 0) and moves the page to the tail of the FIFO queue.

## 2.6 The Least Recently Used (LRU) and Not Frequently Used (NFU Page Replacement Algorithms

A satisfactory estimation to the best algorithm possible relies on our knowledge that pages that have been referenced massively in the near past will apparently be referenced massively again in the foreseeable near future. In contrast, pages that have not been referenced massively for a long time will seemingly remain not referenced for prolonged time. This notion proffers an implementable algorithm: When a page fault happens, evict the page that has not been referenced for the most prolonged time. This idea is called Least Recently Used (LRU) paging.

Despite the fact that LRU is theoretically implementable, it is by no means inexpensive. To completely put LRU into practice, it is essential to keep a data structure – such as a linked list – that holds all of the pages of the memory system, with the Most Recently Used (MRU) page at the beginning and the LRU page at the end. The problem is that the list necessarily has to be updated on each memory reference, all the time. Searching for a page in the list, erasing it, and then transferring it to the beginning of the list is a very time consuming activity, even if it is was implemented in hardware (assuming that such hardware could even be built).

Although there are two LRU algorithms that are implementable in theory [19], there are no machines (except for some small portion of recent computers) that have the appropriate hardware for those algorithms, so there is no concrete benefit in using them. Therefore, there is a need for an algorithm that could be implemented in software. One of the possible algorithms is known as the Not Frequently Used (NFU) algorithm. In this algorithm, a counter is attached to each of the pages, when those counters are initially equal to zero. At every clock interrupt, the OS inspects all the pages in the memory system. For every page the algorithm sums the reference bit (which equal to 0 or 1), and in this way it is possible to get a histogram of how often each of the pages has been referenced. Obviously, when a page fault happens, the page with the smallest counter value is selected for eviction.

The major difficulty with the NFU algorithm is that it is designed to remember all of the information without any option to erase the counters when there is such a need, which consequently may result the OS to evict referenced and crucial pages instead of pages that have been used in the past, but are not used any more. Luckily, a slight improvement to the NFU algorithm enables it to emulate LRU sufficiently well. This improvement of the NFU algorithm is also known as the Aging algorithm [16] [17] [18].

## 2.7 The N-Level Aging Page Replacement Algorithm

The Aging algorithm is a small modification of Not Frequently Used algorithm which makes it possible to simulate Least Recently Used algorithm quite well. Instead of only incrementing the counters of pages referenced, the variation has two parts: First, the counters are shifted right once before the $R$ bit is inserted, meaning that there is actually a division by 2 of the represented decimal number. Second, the $R$ bit is inserted to the leftmost bit, instead of inserting it to the rightmost bit. For instance, if a page has referenced bits 1,1,0,0,0 in the past 5 clock ticks, its referenced counter will look like this: 10000000, 11000000, 01100000, 00110000, 00011000. When a page fault occurs, the page whose counter is the lowest is removed. It is clear that a page that has not been referenced for about K clock ticks will have K leading zeros in its counter (like the referenced counter in the example at the fifth clock tick which has 3 leading zeros after 3 non-referenced clock ticks), and therefore will have a lower value than a counter that has not been referenced for K-1 clock ticks.

The transition of the Aging algorithm to an N-level memory hierarchy model can even add another level of sophistication and optimization, especially because of the existence of a linear proportion between the degradation of the referenced bits and the amount of time that a specific page has not been in use. In this hypothesis, unlike in the other N-level memory hierarchy paging algorithms adaptations and adjustments shown before, there is an interesting phenomenon. Specifically, there is a possibility to create a direct link between the amount of zeros in the beginning of the page referenced bits to the level of memory that that page should be evicted to.

Based on the knowledge that the amount of zeros points to the amount of unreferenced past clock ticks – and therefore on the page aging status – it would be wise to evict the page straight to its proportionate level of memory. Hence, by forming a dynamic pyramid hierarchy of both page and memory necessity it becomes possible to get significantly better performances for a paging algorithm in an N-level memory hierarchy. Therefore, we suggest that the modified Aging paging algorithm will best suit our goal.

*Formulation of the Aging algorithm for N-level memory hierarchy:*

- *Set* memory levels to N (ML = N).
- *Set* current memory level pointer to the highest (L = 1).

1. **Insertion** of a new page:

1.1. *If* the current memory level pointer is higher than the lowest memory level (L > ML) :

1.1.1. *Return* False. /* Recursion Termination */

1.2. **Call** to the page.

1.3. *If* the page exists in the memory / storage:

1.3.1. *Check* if placing in the L-level of memory is possible.

1.3.2. *If* placement possible:

1.3.2.1. *Place* page at the L-level of memory.

1.3.2.2. *Return* True. /* Recursion Termination */

1.3.3. *Else If* placement impossible:

1.3.3.1. *Find* the page with the lowest referenced counter:

1.3.3.1.1. **Remove** the page with the lowest referenced counter.

1.3.3.1.2. *Place* the page instaed of the removed page.

1.3.3.1.3. *Do* **Insertion** of the page with the lowest referenced counter to the proportionate level of memory based on the amount of the zeros in the beginning of the page reference bits (Equation 1).

/* Recursion Invocation */

$$L = \left\lceil \frac{\sum Amount\ of\ Initial\ Zero\ Bits}{\left\lceil \frac{\sum Amount\ of\ Reference\ Bits}{ML} \right\rceil} \right\rceil \tag{1}$$

1.4. *Else If* page does not exist in the memory / storage:

1.4.1. *Return* False. /* Recursion Termination */

2. **Call** to a page:

2.1. *Calculate* the addressing of the page in the memory / storage.

2.2. *If* page found:

2.2.1. *Return* Real Addressing.

2.3. *Else If* page was not found:

2.3.1. *Return* False.

3. **Remove** of a specific page:

3.1. *Store* the page in a temporary storage.

3.2. *Free* the addressing of the page.

3.3. *Return* the page from the temporary storage.

4. **Update** of an existing page (*by the OS*):

4.1. *If* Read / Write action performed on the page:

4.1.1. *Set* R bit to 1 (R = 1).

4.2. *If* clock interrupt:

4.2.1. *Right Shift* one bit to all of the pages counters.

4.2.2. *Add* the R bit to the leftmost bit of all of the pages counters.

# 3 The DeMemory Simulator

## 3.1 Implementation of Memory-Aware Paging Algorithms

As a consequence of our hypothesis regarding the generalization needed to transfer the classic Aging page replacement algorithm to be applicable to multiple levels of memory, the immediate goals are to implement the idea of the Aging algorithm to an N-level memory hierarchy model, and to create a standard simulation for researchers around the world for further research and inspection of the relatively new field of multi-level memory hierarchy. This is especially relevant to the field of Storage Class Memory (SCM), since it is developed rapidly around the globe in the last decade.

Thus, creating a simulator that will be able to run on any computer or sever, and that will be able to simulate a situation in which frames of memory are managed and mapped to specific levels of memory, using slight changes of the original paging algorithms as shown in this thesis, would be beneficial for the development of this technology. To this end, we built the DeMemory simulator framework [20] for the Aging paging algorithm as well as for other algorithms that will be built in the future.

## 3.2 The DeMemory Code Structure

The DeMemory code documentation is presented as processed directly from the Doxygen comments, and it is available in LaTeX and HTML as well [20]. The code is structured as follow, and divided into three main sections:

1. *Classes*: There are 3 different main classes in the code. The first one is Frame, which holds all of the frame information; The second is Algorithm_Data, which holds all of the algorithm data (such as Hits, Misses, page table and victim list); The last one is Algorithm, which contains a pointer to the algorithm data, and includes the name of the specific algorithm, which can be any kind of algorithm, from FIFO to Aging, and a pointer to the algorithm function. In this fashion of dividing the whole structure to those 3 elements we achieve a simple solution in case we need to add a new algorithm to the simulator.

2. *Macros*: The Macros specify the amount of memory levels which is in need. Each of the levels has a macro itself (LEVEL_X) and the number of the level. In this case it is easy to control the levels in a static fashion, which is very suitable for a simulation of memory hierarchies. Also, those macros allow the programmer to change the mapping of the levels easily for testing purposes.

3. *Functions*: The functions are divided into 3 main subsections:

    3.1. *Control functions*: The functions which control the mechanism of the simulator by looping for each page call; getting a random reference when needed; providing all selected algorithms with the random input that was generated; and if there is a need for space adds victimized frame from page table to the list of victims.

    3.2. *Output functions*: The functions which print the help screen, the lists, the detailed statistics, and the final summery of operation.

    3.3. *Algorithm functions*: The functions which are the paging algorithms themselves, including the proxy functions which calculate the direct level to map pages, and the removal and insertion of pages operators.

## 3.3 The DeMemory Simulator Operation

In Linux/UNIX systems, under GCC compiler, and using the Makefile platform, the usage of the DeMemory simulator will be in the following form (Table 1):

>> ./dememory [algorithm] [num_frames] [show_process] [debug] [indexes] [page_refs]

| Parameter | Type | Explanation |
|---|---|---|
| algorithm | {A = Aging,<br>B = N-level Aging} | Page replacement algorithm |
| num_frames | {int > 0} | Amount of page frames in the memory |
| show_process | {1 or 0} | Print page table after each reference is processed |
| debug | {1 or 0} | Debugging output (verbose) |
| indexes | {int > 0} | Amount of unique page indexes |
| page_refs | {int > 0} | Amount of randomized page references |

Table 1: **The DeMemory Parameters, Types and Functionality.**

### 3.4 Example of Usage

→ Run the simulation of the N-level Aging algorithm (which in this DeMemory simulation N is constantly equal to 3 for convenience), with 10 frames in the memory, printing of the page table during the process, without showing debug (verbose), using 100 unique page indexes and 1000 randomized page references.

>> ./dememory B 10 1 0 100 1000

### 3.5 Analysis of Output

The output version, as shown in Figure 6, is based on a simulation of a 3 level memory hierarchy with 10 frames in the memory for each level, when there are 100 unique page indexes and 5 randomized page references. In this case there is a complete printing - both debug and show-process options were selected. Figure 6 presents only the last steps of the simulation.

>> ./dememory B 10 1 1 100 5

```
AGING_N Algorithm
LEVEL: [0] - Frames in Mem: 10, Refs to Mem: 100, Hits: 0, Misses: 4, Hit Ratio: 0.000000, [Max Page calls: 5]
Frame #  :        0        1        2        3        4        5        6        7        8        9
Page Ref :        7       73       72       23        _        _        _        _        _        _
Extra    :  1250000  2500000  5000000        0        0        0        0        0        0        0
Time     : 48296258 48296258 48296258 48296258 48296258 48296258 48296258 48296258 48296258 48296258

Frame #  :        0        1        2        3        4        5        6        7        8        9
Page Ref :        _        _        _        _        _        _        _        _        _        _
Extra    :        0        0        0        0        0        0        0        0        0        0
Time     : 48296258 48296258 48296258 48296258 48296258 48296258 48296258 48296258 48296258 48296258

Frame #  :        0        1        2        3        4        5        6        7        8        9
Page Ref :        _        _        _        _        _        _        _        _        _        _
Extra    :        0        0        0        0        0        0        0        0        0        0
Time     : 48296258 48296258 48296258 48296258 48296258 48296258 48296258 48296258 48296258 48296258

>>>>>>>>>>>>>>> Current Level: [0]

framep->index:[0], framep->extra:[625000], count_zeros_before:[4]
**** calculate_direct_level ****: 1
**** [00001000] ****
INDEX: [0]
INDEXES [0]:[0]
**** REMOVED : [1]****
INSERT-IN ::: framep->index:[0], framep->page:[7]
INSERT ::: framep->index:[0], framep->page:[7]
**** INSERT : [1]****
framep->index:[1], framep->extra:[1250000], count_zeros_before:[3]
framep->index:[2], framep->extra:[2500000], count_zeros_before:[2]
framep->index:[3], framep->extra:[5000000], count_zeros_before:[1]
>>>>>>>>>>>>>>> Current Level: [1]

framep->index:[0], framep->extra:[625000], count_zeros_before:[5]
>>>>>>>>>>>>>>> Current Level: [2]
AGING_N Algorithm
LEVEL: [0] - Frames in Mem: 10, Refs to Mem: 100, Hits: 0, Misses: 5, Hit Ratio: 0.000000, [Max Page calls: 5]
Frame #  :        0        1        2        3        4        5        6        7        8        9
Page Ref :        _       73       72       23       65        _        _        _        _        _
Extra    :        0  1250000  2500000  5000000        0        0        0        0        0        0
Time     :        0 48296258 48296258 48296258 48296258 48296258 48296258 48296258 48296258 48296258

Frame #  :        0        1        2        3        4        5        6        7        8        9
Page Ref :        7        _        _        _        _        _        _        _        _        _
Extra    :   625000        0        0        0        0        0        0        0        0        0
Time     : 48296258 48296258 48296258 48296258 48296258 48296258 48296258 48296258 48296258 48296258

Frame #  :        0        1        2        3        4        5        6        7        8        9
Page Ref :        _        _        _        _        _        _        _        _        _        _
Extra    :        0        0        0        0        0        0        0        0        0        0
Time     : 48296258 48296258 48296258 48296258 48296258 48296258 48296258 48296258 48296258 48296258

AGING_N Algorithm
LEVEL: [0] - Frames in Mem: 10, Refs to Mem: 100, Hits: 0, Misses: 5, Hit Ratio: 0.000000, [Max Page calls: 5]
Elapsed: 0.000706 seconds
```

Figure 6: **DeMemory Example of Output in an N-Level Memory-Aware Aging Paging Algorithm (N = 3).**

As shown in Figure 6, at the beginning the simulation prints the current status of the pages in all of the memory levels. After the N-level Aging algorithm was run on the first level of the memory (*Current Level: [0]*), it was discovered that the first page (*index = 0*) was not referenced for 4 cycles (*extra = 625000, count_zeros_before = 4*) and the algorithm calculated that it should be evicted to a lower level of memory. Therefore, the page has been successfully evicted from the first level of memory (*level = 0 , REMOVED: [1]*) and successfully inserted into the second level of memory (*level = 1, INSERT: [1]*).

Afterwards, the algorithm checked the second and the third levels of memory to determine if there was a need to upgrade or downgrade any of its pages like it has been done in the first level, but concluded there was nothing to do. At the end of the algorithm execution the simulation reprinted the status of the memory and there was an option to see that the actions actually took place.

At the end of the output the program presented the status and the statistics of the simulation so far (and also when the simulation ended), including the number of Hits (access to a page that is found in the memory system) and Misses (access to a page that is not found in the memory system), the total Hit / Miss ratio and the time elapsed ever since the program executed. Those statistics are constantly appended into a log file at the local directory (*dememory.log*).

## 3.6 Algorithms Benchmark

In order to verify our hypothesis regarding the beneficence of using the modified memory-aware Aging page replacement algorithm in multi-level memory hierarchy, and especially when this memory hierarchy is a complex of regular DRAM and different types of Storage Class Memory (SCM), we need to verify that the algorithm is resulting in a more efficient Hit / Miss ratio.

Therefore, we tested and compared the two types of algorithms, while the first was running on a classic one-level memory (DRAM only), and the second was running on a 3-level hierarchy as follows: a classic one-level memory and two extra memory levels with the same volume (for accurate comparability measurements) as the DRAM. Those two extra levels were simulating two different types of SCM – one which was 2 times slower than the DRAM, and the other which was 3 times slower than the DRAM.

These architectures created a situation in which while the classic memory hierarchy volume is $C$ with $V$ speed of reference, the complex 3-level memory hierarchy volume is $3C$ with $0.5V$ speed of reference on average. The explanation of those parameters is simple: Each addition of a memory level to the memory hierarchy ($N$) adds to the capacity of the whole memory complex (*Volume*), but also slows the memory complex in average (*Speed*). In this example that is the reason why the volume of the memory becomes 3 times bigger (1+1+1 = 3) but also 2 times slower ((1+2+3)/3 = 2). Those calculations can be shown in an equation form as the following (Equation 2, 3):

$$Volume_{Total}=\sum_{i=1}^{N} C_i \tag{2}$$

$$Speed_{Total}=\frac{\sum_{i=1}^{N} V_i}{N} \tag{3}$$

Therefore, it was crucial to verify that although some slowdown has occurred, in various cases the N-level memory-aware Aging paging algorithm still delivers better performance than the classic Aging paging algorithm (with no extra memory levels and without memory-awareness).

## 3.7 Results and Analysis

As previously mentioned, the benchmark of the DeMemory simulator as shown in this thesis has been performed on two different architectures, using several parameters. The following graphs show this benchmark result, the Hit / Miss ratio, which will be presented as a function of three variables: The amount of frames in memory (F), the amount of unique page indexes (I) and the amount of page references (R). In each benchmark we set two of those parameters to be fixed, and ranged the third parameter in two scales: The first scale ranged from 10 till 100 (Figure 7-9), and the second scale ranged from 1000 to 1 million (Figure 10-12). The purpose of those two scales is to examine the performance of the simulation in normal usage scale and in High Performance Computing (HPC) scale.

The results are as followed:

- The Hit / Miss ratio as function of the amount of frames in memory (Between 10 and 100) shows that although the N-level memory-aware Aging paging algorithm creates a better Hit / Miss ratio when the number of frames is low, as frames added to the memory system there is an advantage to the classic 1-level Aging paging algorithm (Figure 7). Those results also persist when the range of the amount of frames in the memory rise to HPC levels (1e3 – 1e6), then the Hit / Miss ratio becomes constant with a clear favor of the classic 1-level Aging paging algorithm (Figure 10). However, although these results seems to present an advantage of the classic 1-level Aging paging algorithm, it is worth noticing that it is impractical to enlarge the DRAM to this kind of capacity because of its high cost – the price gap between Storage Class Memory (SCM) and DRAM is about an order of magnitude – meaning that it would be cost-ineffective, and that the Hit / Miss ratio gap between the two algorithms (~10%) is not big enough to justify that cost.

- The Hit / Miss ratio as function of the amount of unique page indexes (Between 10 and 100) shows that although the classic 1-level Aging paging algorithm creates a better Hit / Miss ratio when the number of unique page indexes is the same as the amount of frames in the memory (a situation that almost never happens), as unique page indexes added to the system there is a clear advantage to the N-level memory-aware Aging paging algorithm (Figure 8), as it results Hit / Miss ratio which is N times better than the classic 1-level Aging paging algorithm (in this case, 3 times better). Thus we can conclude that there is at least a linear proportion between the amount of extra memory levels and the Hit / Miss ratio while using the N-level memory-aware Aging paging algorithm. These results also persist when the range of the amount of unique page indexes in the memory rises to HPC levels (1e3 – 1e6), which then the Hit / Miss ratio becomes constant with clear favor of the N-level memory-aware Aging paging algorithm, as the classic 1-level Aging paging algorithm fails to deliver an acceptable Hit / Miss ratio (Figure 11), meaning the gap is N times better at the least, and infinity better at the most.
- The Hit / Miss ratio as a function of the amount of page references (Between 10 and 100) shows that only when the amount of page references is equal or less to the amount of frames in the memory the two algorithms results the same Hit / Miss ratio, which is equal to zero (because every page insertion is a miss at the beginning). Yet, when the amount of page references gets bigger than the amount of frames in the memory there is a clear advantage to the N-level memory-aware Aging paging algorithm (Figure 9), as it results Hit / Miss ratio which is N times better than the classic 1-level Aging paging algorithm (in this case, 3 times better). These results also persist when the range of the amount of page references in the memory rises to HPC levels (1e3 – 1e6), which then the Hit / Miss ratio becomes constant (Figure 12) and stays N times better than the classic 1-level Aging paging algorithm (in this case, 3 times better). Thus we can conclude that there is a clear linear proportion between the amount of extra memory levels and the Hit / Miss ratio while using the N-level memory-aware Aging paging algorithm.

## 3.8 Comparison of Different N-Level Algorithms

In addition to the previous results, there was a need to examine and compare different algorithms for the same system of N-levels, for a fixed N. To this end, we examined the results of the N-level memory-aware Aging paging algorithm with intentional modification where there is no direct link between the amount of zeros in the beginning of the page referenced bits to the level of memory that that page should be evicted to.

As mentioned in chapter 2.7, based on the knowledge that the amount of zeros in the beginning of the page referenced bits points to the amount of unreferenced past clock ticks – and therefore on the page aging status – it was wise to evict the page straight to its proportionate level of memory. In order to test the hypothesis that by forming a dynamic pyramid hierarchy of both page and memory necessity it was possible to get the best performances for a paging algorithm in an N-level memory hierarchy, we modify the behavior of the algorithm to select different levels rather than the correct direct levels. Specifically, if a page at level 1 was directed towards level 2 in our original N-level algorithm, it is actually redirected to level 3 and vice versa. Afterwards, we reexamined the Hit / Miss ratio as function of the amount of page references (Figure 13) and discovered, unsurprisingly, that there was a loss in performance in comparison to the correct algorithm.

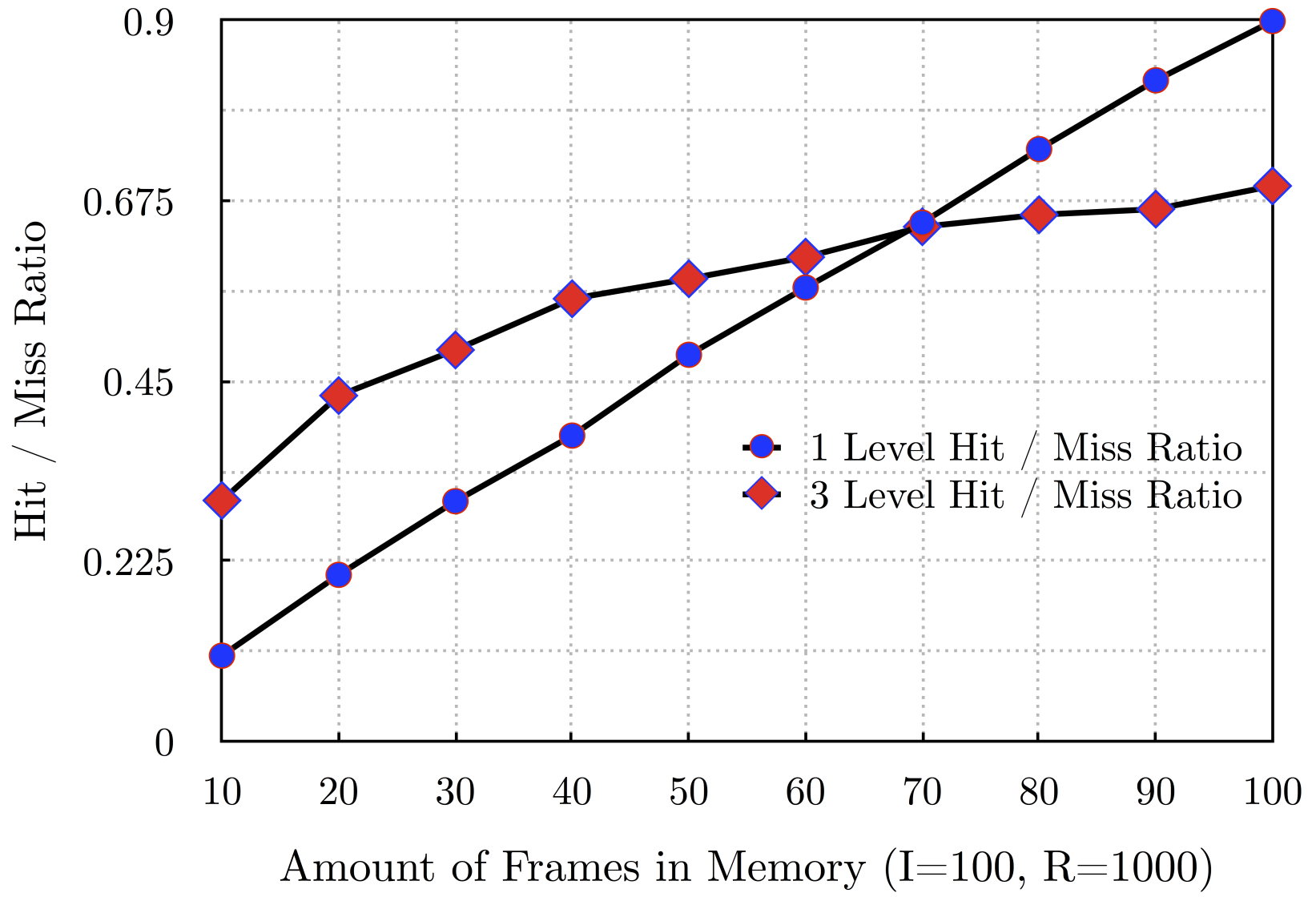

Figure 7: **The Hit / Miss Ratio as Function of the Amount of Frames in Memory (F).**

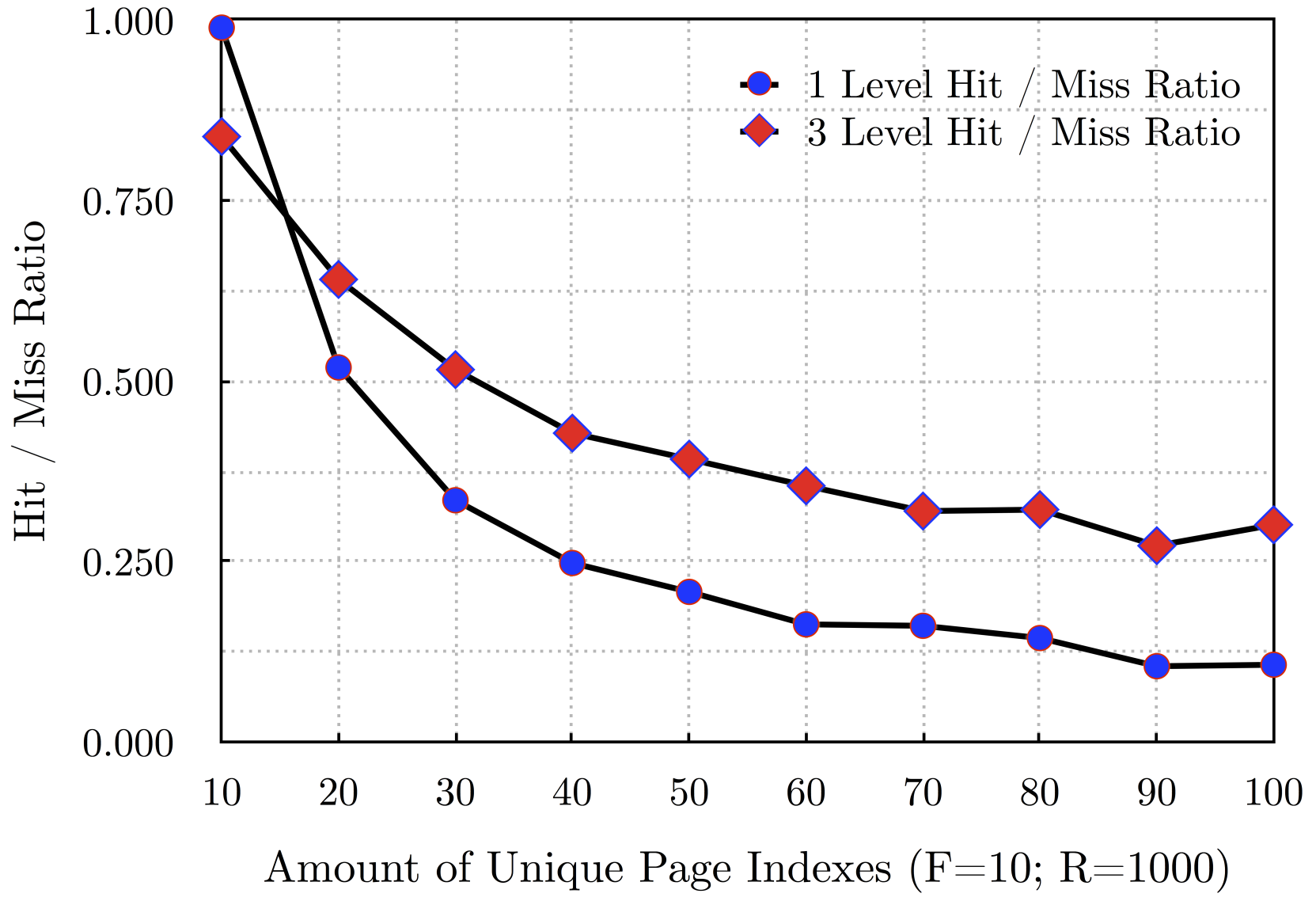

Figure 8: **The Hit / Miss Ratio as Function of the Amount of Unique Page Indexes (I).**

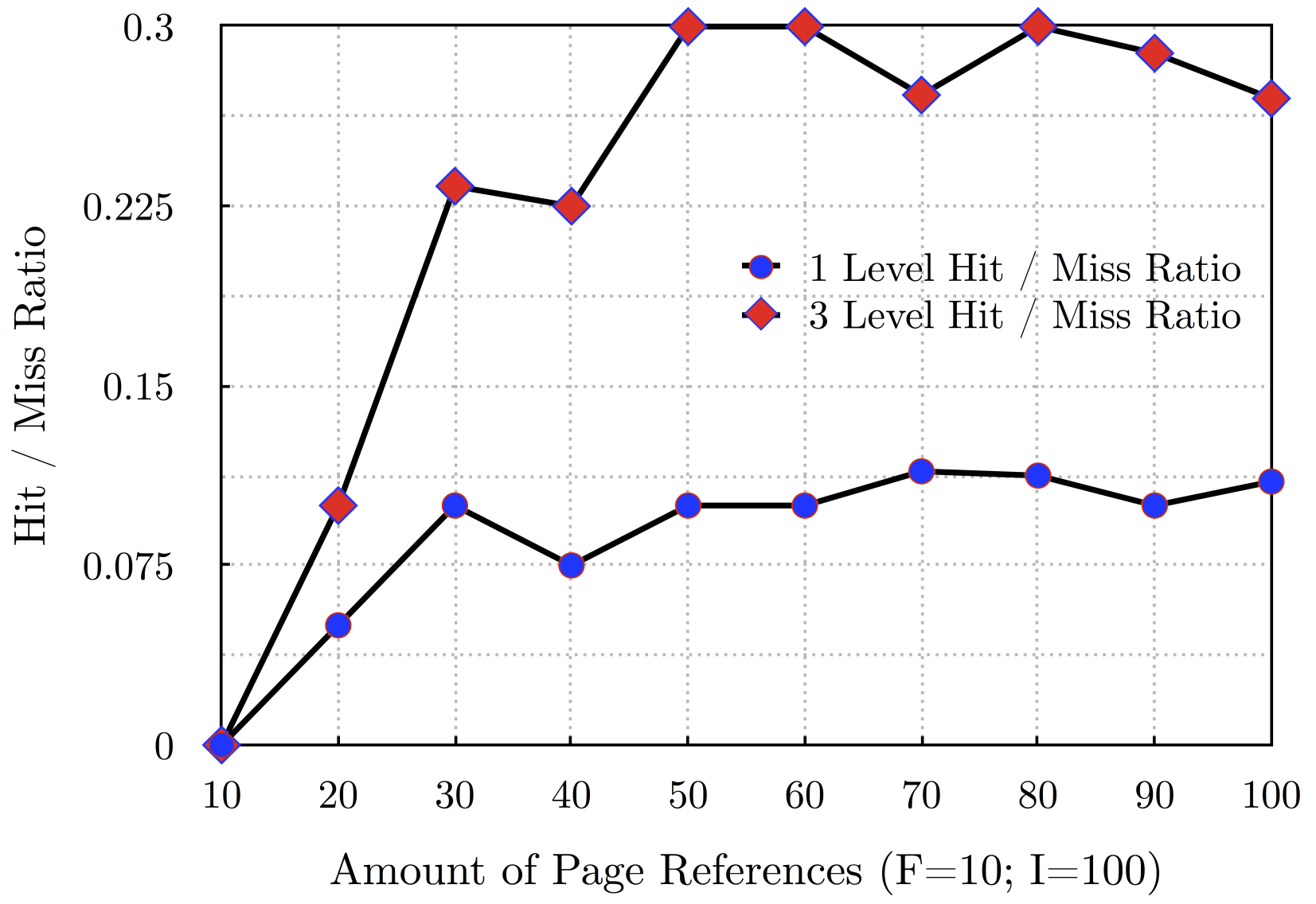

Figure 9: **The Hit / Miss Ratio as Function of the Amount of Page References (R).**

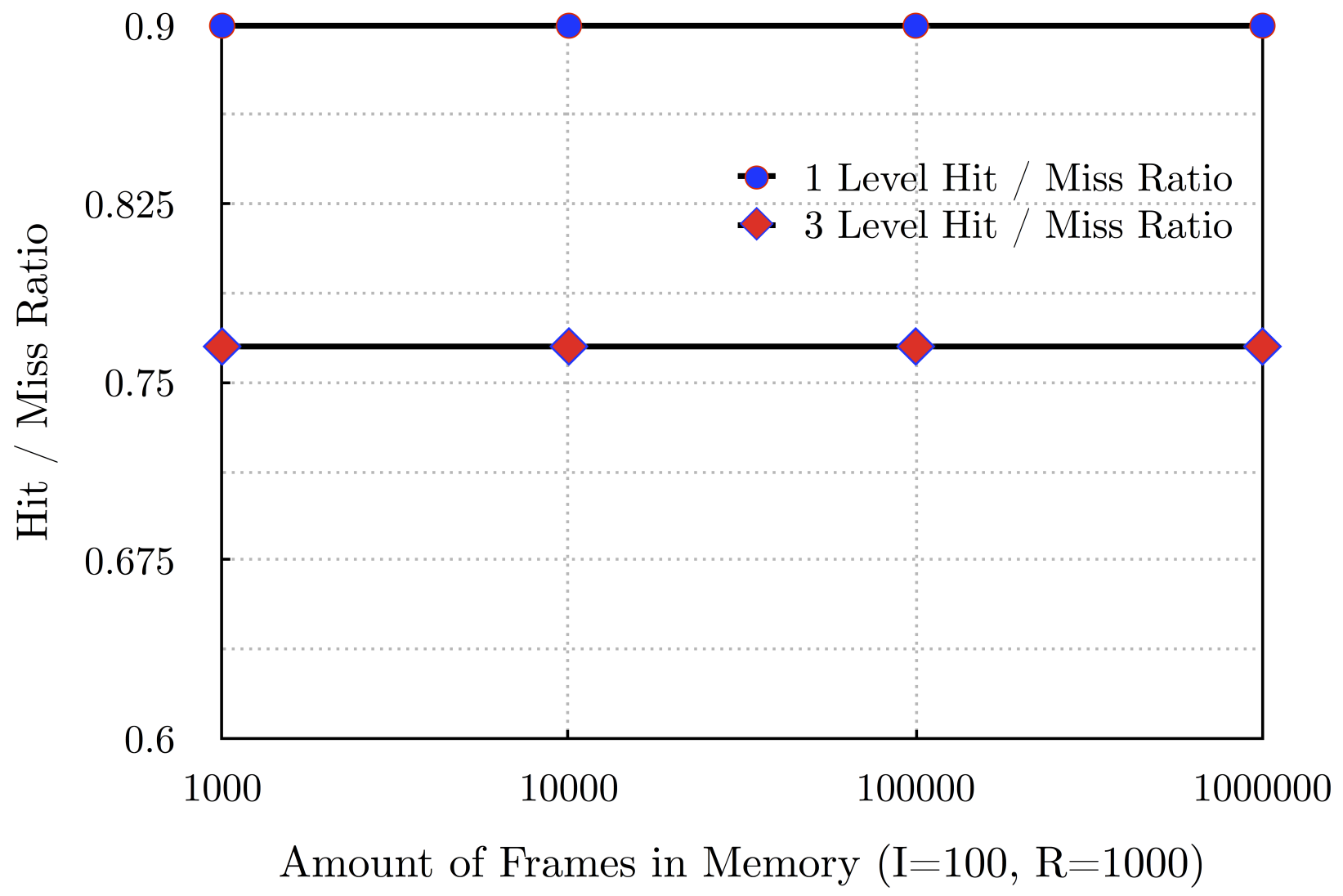

Figure 10: **The HPC Hit / Miss Ratio as Function of the Amount of Frames in Memory.**

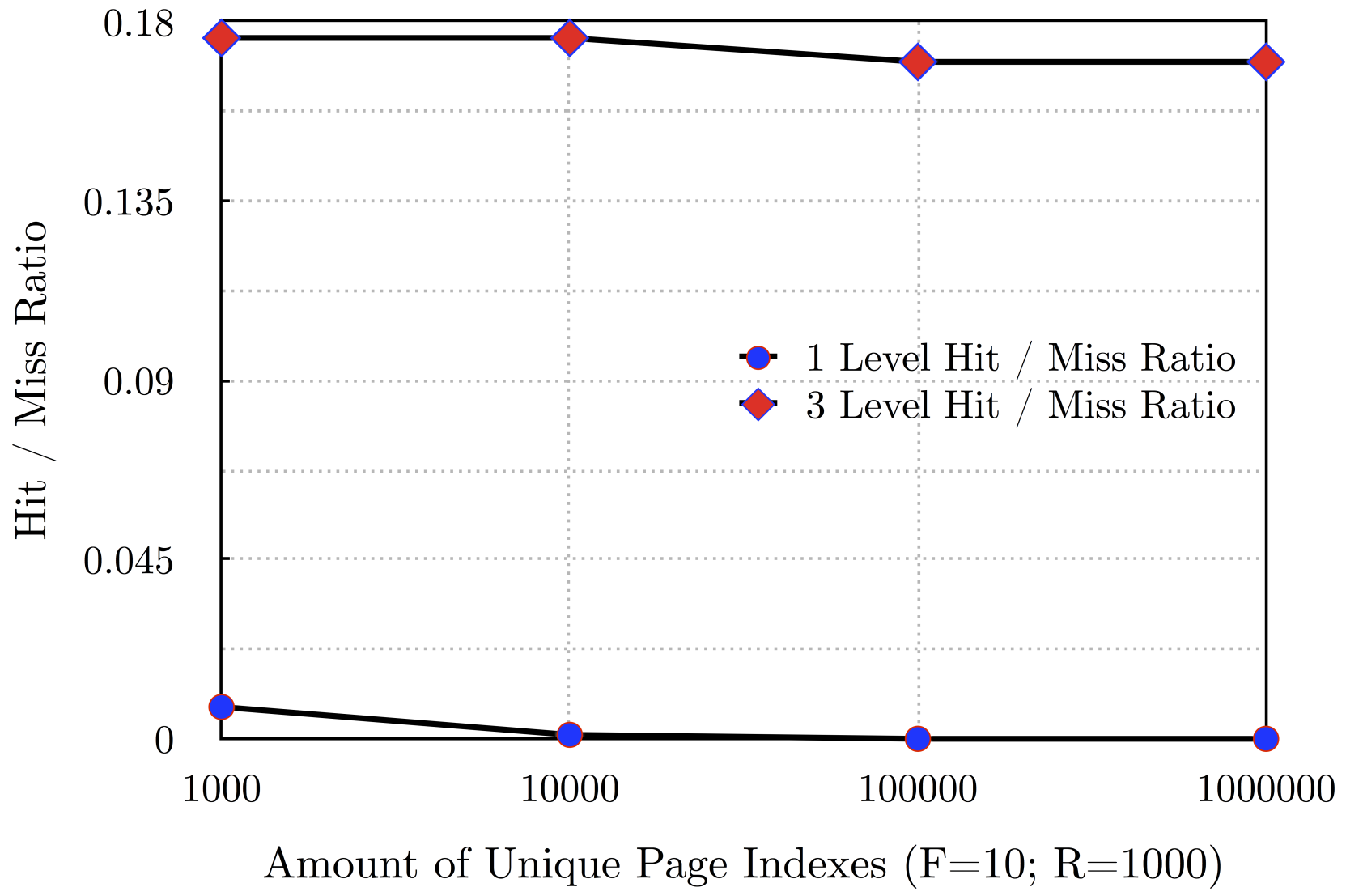

Figure 11: **The HPC Hit / Miss Ratio as Function of the Amount of Unique Page Indexes.**

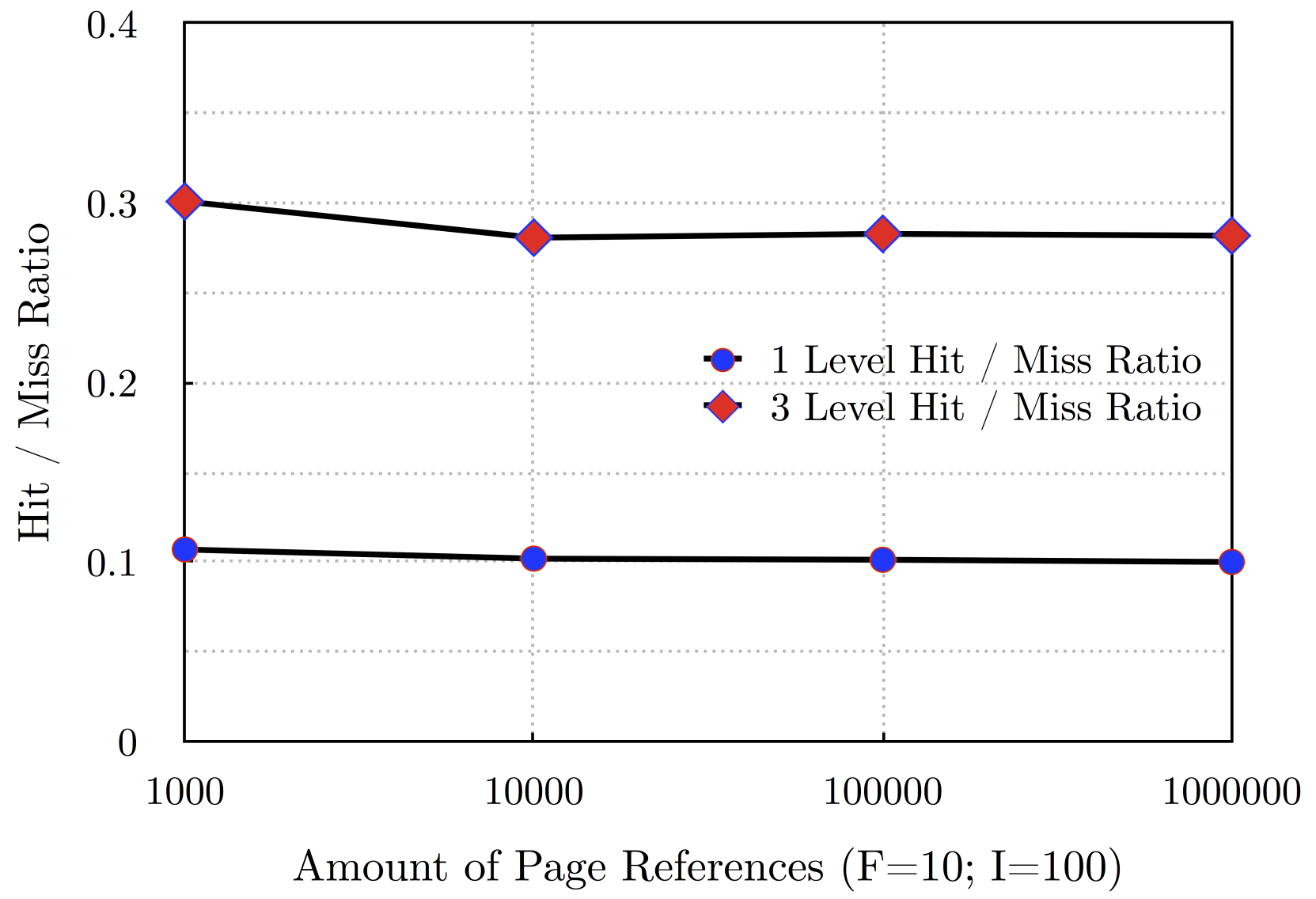

Figure 12: **The HPC Hit / Miss Ratio as Function of the Amount of Page References.**

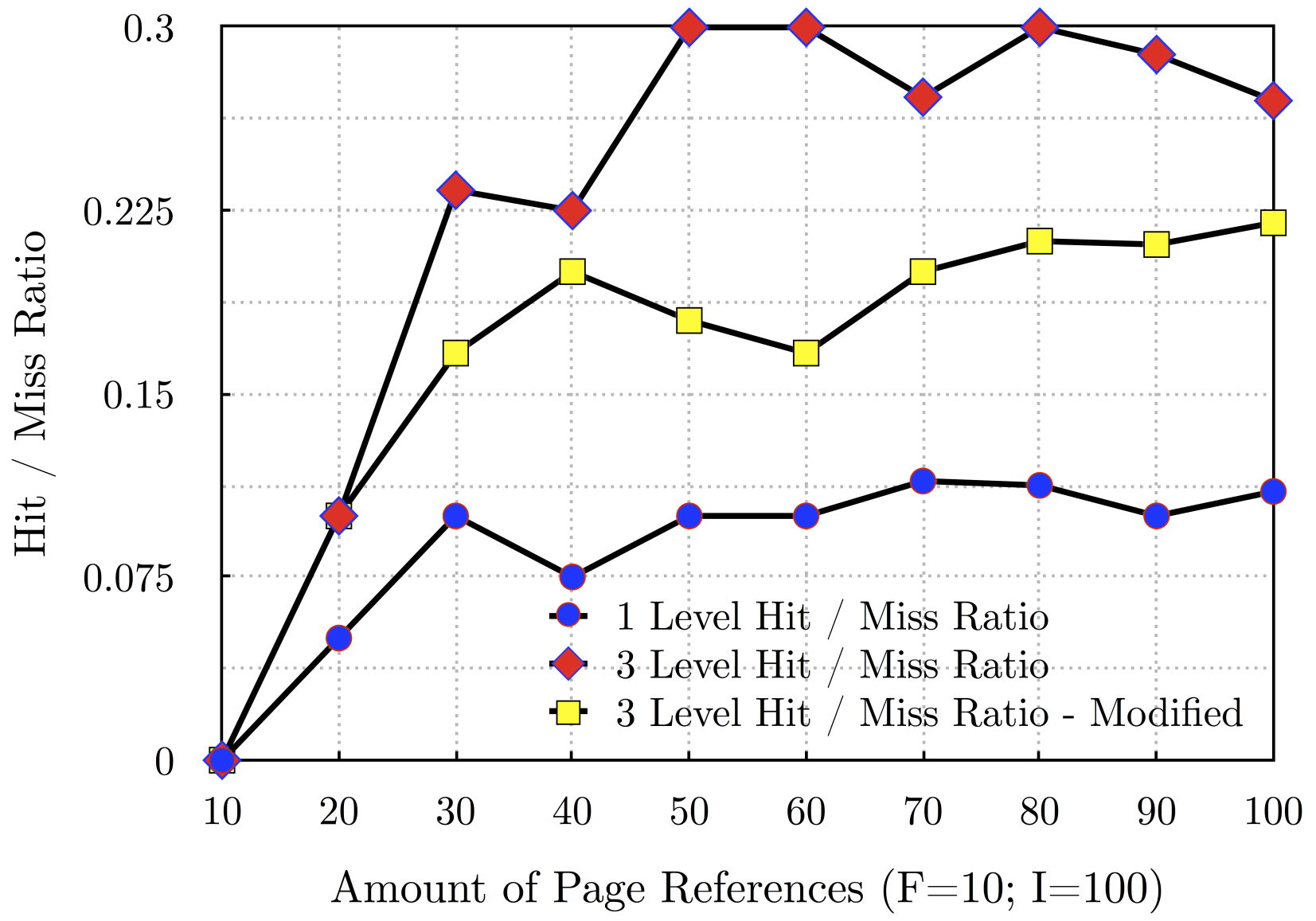

Figure 13: **The Hit / Miss Ratio as Function of the Amount of Page References including the Hit / Miss Ratio for a Modified Version of the N-Level Memory-Aware Aging Paging Algorithm.**

# 4 Conclusions and Future Work

## 4.1 Conclusions

Those benchmarks, results and analysis lead to the following conclusions:

1. In average, a complex of regular DRAM and different types of Storage Class Memory (SCM) with the same size, will make the whole memory about N times slower than the DRAM, but N times bigger than the DRAM.

2. There is a clear advantage to the N-level memory-aware Aging paging algorithm, as it results Hit / Miss ratio which is N times better than the classic 1-level Aging paging algorithm.

Therefore, because these two conclusions at least eliminate each others effect by compensating the slowdown with at least the same growth of the Hit / Miss ratio, and because of the fact that the price gap between SCM and DRAM is about an order of magnitude, we can conclude that it would be beneficial to use the N-level memory-aware Aging paging algorithm in multi-level memory hierarchy which include SCM in regular computing systems as well as in High Performance Computing clusters.

## 4.2 Future work

This thesis opens a number of prospective directions for future research. One immediate direction is to explore how the current N-level memory-aware Aging paging algorithm is reacting when the memory levels are not in the same size or are not from the same class of the current Storage Class Memories (SCM) which are in the market.

Finally, we also expect that in the near future the SCM will be a real and widespread technology, meaning that investigating the real technology and comparing it to the current simulation would be a fertile ground for further research and development.

# תקציר

זיכרונות המחשב – הזיכרון הראשי וזיכרון האחסון – נתפשים עד היום כמשאבים חד-ממדים ובלתי-מתוחכמים, עם מספר תכונות פשוטות, כמו זמן גישה קבוע. בחלוף השנים תפישה זו הוכחה כשגוייה בתכלית, וכיום הבנה של מדרג הזיכרון יכולה להיות שימושית במטרה לשפר את הביצועים של אלגוריתם או מבנה נתונים. לדוגמה, ה-Storage Class-Memory (או ה-SCM), הינו סוג של טכנולוגיה חדשה אשר מייצגת צורת כלאיים בין הזיכרון הראשי וזיכרון האחסון, עם מאפיינים ייחודיים כך שזיכרון זה הינו בלתי-נדיף, זול במונחי עלות לסיבית, בעל זמני גישה מהירים הן עבור כתיבה והן עבור קריאה, וכן הינו זיכרון אשר אינו מכיל חלקים נעים.

מערכות הפעלה עתידות להשתמש ב-SCM כמכשירי זיכרון-אחסון מהירים מאוד המעוצבים על ידי מערכת קבצים ומבני-נתונים, או לחילופין כקבצי זיכרון ראשיים עבור הדור הבא של תוכנות המחשב. בעתיד הנראה לעין ה-SCM מתעתד לשנות את צורת תוכנות המחשב החדשות, את אופן הגישה לזכרון האחסון, ואת הדרך בה זיכרונות אחסון נבנים. על כן, שילוב בין ה-SCM ובין מנהל הקצאות זיכרון ייעודי (Memory Allocation Manager – MAM) אשר יאפשר למתכנת לשלוט באופן ידני בזיכרונות השונים שבמדרג הזיכרון, מתעתד להשיג רמה חדשה של ביצועים עבור מבני-נתונים בעלי מודעות למדרג הזיכרון. עם זאת, על אף ששליטה ידנית ב-MAM נדמית כגישה האופטימלית לניהול מדרג זיכרון עם זיכרונות רבים, טכניקה זו עודה רחוקה מאוד מלהיות ממשית, והסיכויים שהיא תמומש עבור הקודים הנוכחיים המשתמשים בפלטפורמות מחשוב עתיר ביצועים (High Performance Computing – HPC) הינם נמוכים למדי.

משמעות הנחת מוצא זו היא שהדרך המתקבלת ביותר על הדעת להציג את ה-SCM לתוך מערכת זיכרון ממשית תהיה באמצעות מימוש גרסה אוטומאטית של ה-MAM, תוך היעזרות בעקרונות היסוד של אלגוריתמי הדפדוף המוכרים עבור מדרג זיכרון סטנדרטי. הנחת היסוד שלנו היא שהשגת עֲבִירוּת ראויה בין רמות הזיכרון הינה בגדר האפשר תוך שימוש ברעיונות האלגוריתמים המיושמים כיום במערכות הזיכרון הווירטואלי, וכן שההתאמה של אלגוריתמים אלו ממדרג זיכרון סטנדרטי למדרג זיכרון עם ריבוי רמות זיכרון הינה אף היא בגדר האפשר.

במטרה להראות את הנחות היסוד שלנו חקרנו מספר אלגוריתמי דפדוף, ומתוכם סיננו את אלו שיכולים לעבור התאמה מוצלחת ממדרג זיכרון סטנדרטי למדרג זיכרון מרובה רמות זיכרון. במהלך המחקר גילינו כי שימוש בהתאמה מסויימת של אלגוריתם הדפדוף הידוע כ-Aging למדרג זיכרון מרובה רמות זיכרון מניב את הביצועים הטובים ביותר במונחי "יחס פגיעה-החמצה". במטרה לאשרר השערות אלו בנינו מַדְמֶה (העונה לשם "ה-DeMemory") על מנת לנתח את ביצועי האלגוריתמים שלנו, וכן על מנת לנתח באמצעותו אלגוריתמים נוספים שיתוכננו בעתיד.

# תוכן עניינים

האוניברסיטה הפתוחה

המחלקה למתמטיקה ומדעי המחשב

# ייעולים באלגוריתמי ניהול

# עבור מדרג זיכרון מרובה רמות

עבודה זו הוגשה כחלק מהדרישות לקבלת תואר

"מוסמך במדעים" M.Sc. במדעי המחשב

באוניברסיטה הפתוחה

החטיבה למדעי המחשב

על-ידי

**גל אורן**

דצמבר 2015